\documentclass[%
 reprint,
 superscriptaddress,
onecolumn,
nofootinbib,
 amsmath,amssymb,
 aps,
 pra,
floatfix,
]{revtex4-2}

\usepackage{graphicx}
\usepackage{dcolumn}
\usepackage{bm}
\usepackage{xcolor}
\usepackage{hyperref}
\usepackage{placeins}
\usepackage[mathlines]{lineno}
\usepackage{setspace}

\newcommand{\subhead}[1]{\textbf{#1}\hspace{1em}}

\newcommand{\stressunit}[0]{\epsilon\sigma^{-3}}
\newcommand{\fluxunit}[0]{(m\epsilon)^{1/2}\sigma^{-3}}
\newcommand{\velunit}[0]{(\epsilon/m)^{1/2}}
\newcommand{\densunit}[0]{m\sigma^{-3}}

\begin{document}

\title{Active learning for parameter-free multiscale modeling of boundary lubrication}

\author{Hannes Holey}
\email{hannes.holey@unimi.it}
\affiliation{%
Department of Physics, Center for Complexity and Biosystems, University of Milan, Via Celoria 16, 20133 Milan, Italy
}
\affiliation{%
 Institute for Applied Materials, Karlsruhe Institute of Technology, Straße am Forum 7, 76131 Karlsruhe, Germany
}%
\affiliation{%
Department of Microsystems Engineering (IMTEK), University of Freiburg, Georges-Köhler-Allee 103, 79110 Freiburg, Germany
}%

\author{Peter Gumbsch}
\affiliation{%
 Institute for Applied Materials, Karlsruhe Institute of Technology, Straße am Forum 7, 76131 Karlsruhe, Germany
}%
\affiliation{%
Fraunhofer Institute for Mechanics of Materials IWM, Wöhlerstraße 11, 79108 Freiburg, Germany
}%

\author{Lars Pastewka}
\email{lars.pastewka@imtek.uni-freiburg.de}
\affiliation{%
Department of Microsystems Engineering (IMTEK), University of Freiburg, Georges-Köhler-Allee 103, 79110 Freiburg, Germany
}%
\affiliation{%
Cluster of Excellence livMatS, Freiburg Center for Interactive Materials and Bioinspired Technologies, University of Freiburg, Georges-Koehler-Allee 105, 79110 Freiburg, Germany
}

\date{\today}

\begin{abstract}
Lubricated friction is a multiscale problem where molecular processes dictate the macroscopic response of the system.
Traditional lubrication models rely on semi-empirical constitutive relations, which become unreliable under extreme conditions.
Here, we present a simulation framework that seamlessly couples molecular and continuum models for boundary lubrication without fixed-form constitutive laws.
We train Gaussian process regression models as surrogates for predicting interfacial shear and normal stress in molecular dynamics simulations.
An active learning algorithm ensures that our model adapts in scenarios where common constitutive laws fail, such as near phase transitions.
We demonstrate our approach for nanoscale fluid flow over rough and heterogeneous surfaces, paving the way for accurate boundary lubrication simulations at experimental length and time scales.
\end{abstract}

\maketitle

\section*{Introduction}

Many phenomena in nature and engineering are multiscale: Molecular processes interact with larger-scale features to determine the measurable, real-world response of the system.
An important example is friction, where interfacial slip is controlled by a combination of molecular interactions across the interface~\citep{israelachvili1991_intermolecular}, the geometric features of surface roughness~\citep{sayles1978_surface}, and the elastic energy stored in the structural deformation of the contacting bodies~\citep{bayart2016_slippery}.
Such mechanical processes cause earthquakes~\citep{kanamori2004_physics} and are responsible for a large fraction of energy dissipated in machinery~\citep{holmberg2017_influence}.
While the construction of numerical multiscale models for fracture~\citep{kohlhoff1991_crack,kermode2008_lowspeed} and plasticity~\citep{tadmor1996_quasicontinuum,shilkrot2004_multiscale} has been carried out for at least three decades, modeling friction has relied on combining insights from single-scale molecular or continuum calculations, limiting the predictive power of numerical models.
We here present a multiscale simulation framework that uses nonparametric statistical models from machine learning to pass information between molecular and continuum scales.
Our results demonstrate the concurrent multiscale modeling of lubricated frictional interfaces, enabling the design of tribological systems.
The model does not require the specification of fixed-form constitutive relations for the response of the liquid, as commonly employed in lubricant design~\citep{jadhao2017_probing,falk2020_nonempirical}.
We show that this captures even situations where no constitutive model is applicable, such as for lubricants that are pressurized while geometrically confined between contacting peaks on the rough topography and undergo jamming transitions from liquid to solid states.

The role of viscous lubricant films is to separate the two sliding surfaces.
Friction reduces because dissipation occurs predominantly within the film.
Driven by the need to reduce energy dissipation, tribological research is currently moving towards low-viscosity lubricants \citep{zhang2016_recent}.
This reduces the film thickness toward mixed and boundary lubrication regimes, where the lubricating film is not much thicker than the average height of peaks on the contacting rough topographies, also called asperities.
Lubricants then experience extreme conditions in some regions of the contacting bodies, where fluid films have a thickness on the order of the size of the lubricant molecules \citep{archard1962_lubrication,glovnea2003_measurement}, while other regions remain well-lubricated.
Molecular specificity can no longer be neglected at these scales, controlling effects like ordering~\citep{thompson1990_shear}, density layering~\citep{gao1997_layering}, solvation forces~\citep{horn1981_direct}, and fluid-wall slip~\citep{pit2000_direct,zhu2001_ratedependent,cheng2002_fluid}.
The small regions of intimate contact then dominate the macroscopic friction measured in experiments.

Atomistic modeling techniques have become a standard tool in tribology~\citep{ewen2018_advances},
providing mechanistic insight into the energy dissipation and mechanochemistry of dry and lubricated contacts~\citep{pastewka2010_atomistic,kuwahara2019_mechanochemical}.
Yet, incorporating these atomistic insights into large scale models remains challenging.
So far, most approaches are of sequential nature, where atomistic simulations parametrize constitutive laws that are subsequently used on the continuum scale \citep{martini2006_molecular,savio2015_multiscale,savio2016_boundary,codrignani2023_continuum}. 
With carefully chosen constitutive models, these approaches successfully bridge scales and are of particular use when experimental calibration is challenging.
However, by construction, sequential models do not account for the feedback from the large to the small scale---an aspect that is important in tribology, since dynamic frictional systems may evolve into regions not considered for calibration and minuscule changes in surface inclination may lead to significant changes in friction~\citep{yue2024_influence}.

Concurrent multiscale schemes solve the small scale and the large scale problem at the same time, enabling the two-way coupling missing in sequential models.
One class of concurrent multiscale schemes has been unified under the umbrella term \emph{heterogeneous multiscale methods} (HMM) \citep{e2003_heterogeneous,e2003_heterogeneousa}.
The key idea is to solve a macro scale problem, where missing data comes from representative microscale simulations, which are in turn constrained by the macro solution.
For instance, in grid-based continuum solvers for the Navier-Stokes equation, molecular dynamics (MD) simulations provide time-averaged stress tensor components at each grid point given its local strain rate.
This is possible when the small scale problem converges to a steady state much faster than the macroscopic variables change, i.e., in the case of large timescale separation, a key assumption of HMM.
The HMM and various flavors thereof have been applied to small scale flow problems with atomistic-continuum coupling
\citep{ren2005_heterogeneous,yasuda2008_model,asproulis2013_artificial,borg2013_fluid,borg2013_multiscale,borg2015_hybrid,stalter2018_molecular,tedeschi2022_multiscale}.
Among the drawbacks of HMM are the enslavement to the smallest time step and the handling of noisy data on the continuum scale, which is why many authors fall back to precomputed lookup tables and interpolation using surrogate models \citep{asproulis2013_artificial,stalter2018_molecular,tedeschi2022_multiscale}.
This leads to probabilistic machine learning techniques, such as Gaussian process (GP) regression~\citep{rasmussen2006_gaussian}, which can handle measurement noise and yield measures of the error made when interpolating.

GP surrogates can make fast and accurate predictions for high-dimensional inputs and have recently been successful to predict quantum mechanical energies and forces based on atomic environments~\citep{bartok2010_gaussian}.
Probabilistic surrogate models have also been used in continuum solid mechanics, e.g., to couple crystal plasticity with coarse-scale continuum calculations \citep{knap2008_adaptive}, for one-dimensional heat transfer \citep{salloum2012_stochastic}, or to study the deformation of microstructured materials \citep{rocha2021_onthefly}.
The GP uncertainty allows constructing so-called \emph{on-the-fly} schemes \citep{rocha2021_onthefly,li2015_molecular,stephenson2018_accelerating,zhao2018_active}, that update the underlying training database during a simulation, if new inputs are not well represented by the initial training data.
This is particularly useful for problems with sudden changes in the constitutive behavior, for instance due to phase transformations or chemical reactions.
Frameworks in which the GP uncertainty guides not only the ``when'' but also the ``where'', i.e., the locations (in input or feature space) where new training data is acquired, are historically referred to as kriging~\citep{krige1951_statistical}, or nowadays, active learning.

In this work, we present a concurrent multiscale framework that couples molecular and continuum simulations of lubricant flow (see Methods).
On the continuum scale, we solve the mass and momentum balance for thin film flows, where we leverage the thin film assumption to average out the gap coordinate \citep{holey2022_heightaveraged}.
In contrast to conventional reduced-order equations in lubrication \citep{reynolds1886_iv}, our scheme is agnostic of the functional form of the constitutive law.
We achieve this by using GP regression to construct surrogate models directly for the stress tensor, informed by non-equilibrium MD simulations of highly confined fluids. 
The surrogate avoids redundant and expensive MD simulations by interpolating between prior calculations.
We use the posterior variance of the GP to determine where new molecular simulations are necessary, leading to an active learning or ``learn-on-the-fly'' scheme for multiscale lubrication.

\section*{Results}

\begin{figure*}[!ht]
\includegraphics[width=\textwidth]{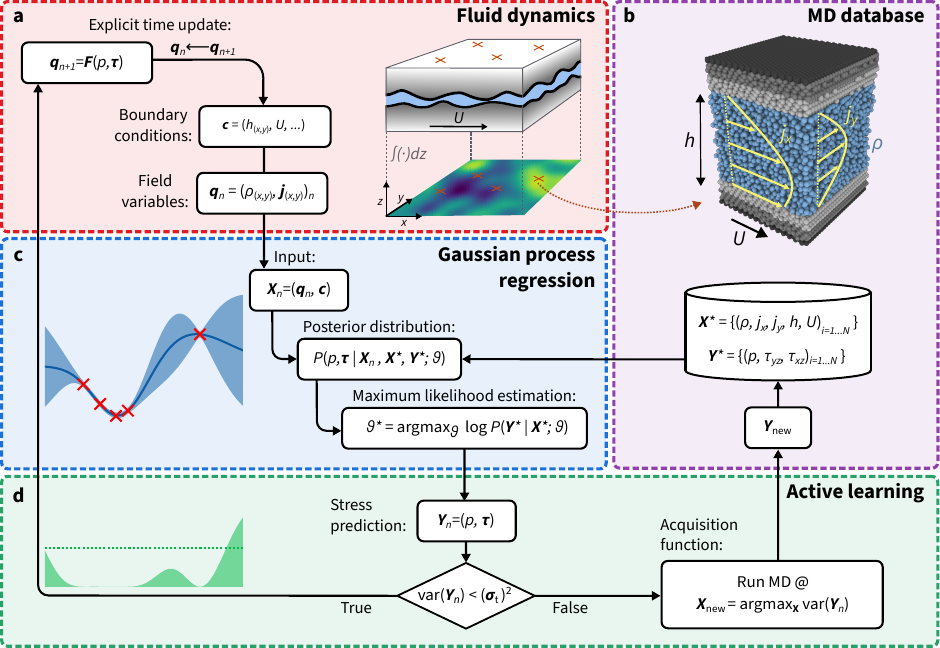}
\caption{%
\textbf{Graphical representation of the active learning workflow for nanoscale lubrication problems.}
\textbf{a} At every time step $n$, as long as the continuum solver has not converged to a steady state solution, we calculate the explicit time update of the unknown field $\mathbf{q}_{n+1}$.
The stress components that determine the update depend on the field variable at the current time increment $\mathbf{q}_n$, and on constant boundary conditions $\mathbf{c}$, such as the gap height distribution $h(x,y)$ and the wall velocity $U$.
Normal ($p$) and shear stress components $\bm{\tau} = (\tau_{xz}, \tau_{yz})^\top$ are calculated with non-equilibrium molecular dynamics (MD) simulations  (\textbf{b}).
Field variables and boundary conditions span the input space $\mathbf{X}_n$ of our stress surrogates, build as Gaussian process (GP) regression models (\textbf{c}).
In the two-dimensional case, we consider three separate GP models, one for each stress component.
The shear stress models are multi-output GPs, which predict the shear stress at the top and the bottom surface with a common kernel.
Hyperparameters are optimized using maximum likelihood estimation (MLE) every time new data is added to the training database.
\textbf{d} The stress prediction is evaluated based on the GP posterior variance.
If the maximum variance is larger than a prescribed tolerance $\sigma_\mathrm{t}$, additional MD data is acquired.
Otherwise, the macro solver proceeds with a time update.}
\label{fig:workflow}
\end{figure*}

\subhead{Coupling framework}
Figure~\ref{fig:workflow} illustrates the coupling between continuum fluid dynamics and molecular dynamics simulations, enabled by GP regression and an adaptively updated database using active learning.
The continuum solver explicitly updates mass and momentum densities according to gap-averaged conservation laws, which requires knowledge of the interfacial stress state (see Methods).
Next to the density fields, the gap topography and the sliding velocity determine the stress (see Fig.~\ref{fig:workflow}a).
Thus, for a given point in the quasi two-dimensional domain, we calculate the normal and shear stress components with non-equilibrium MD simulations (see Fig.~\ref{fig:workflow}b).
Since running an MD simulation at each grid point of the continuum scheme would be computationally unfeasible, we interpolate between the results of only a few MD runs in the input space spanned by the continuum field variables, the topography, and the boundary conditions (such as the sliding velocity).
Our surrogate model based on GP regression provides stress predictions as well as their uncertainties (see Fig.~\ref{fig:workflow}c).
The training database successively grows during the course of a simulation by the addition of new MD results each time the prediction uncertainty does not meet a pre-defined criterion.
Every addition of training data leads to a readjustment of the GP's hyperparameters by maximizing the likelihood of the posterior distribution (see Methods).
The location of new training data is determined by an active learning framework (see Fig.~\ref{fig:workflow}d), which requests new MD runs under conditions that show the largest uncertainty at the current time step.
Our multiscale framework, consisting of the continuum solver, the orchestration of MD simulations, and GP regression, has been made publicly available under the terms of the MIT license\footnote{\url{https://github.com/hannes-holey/hans}}.

\begin{figure*}[!ht]
\centering
\includegraphics[width=\textwidth]{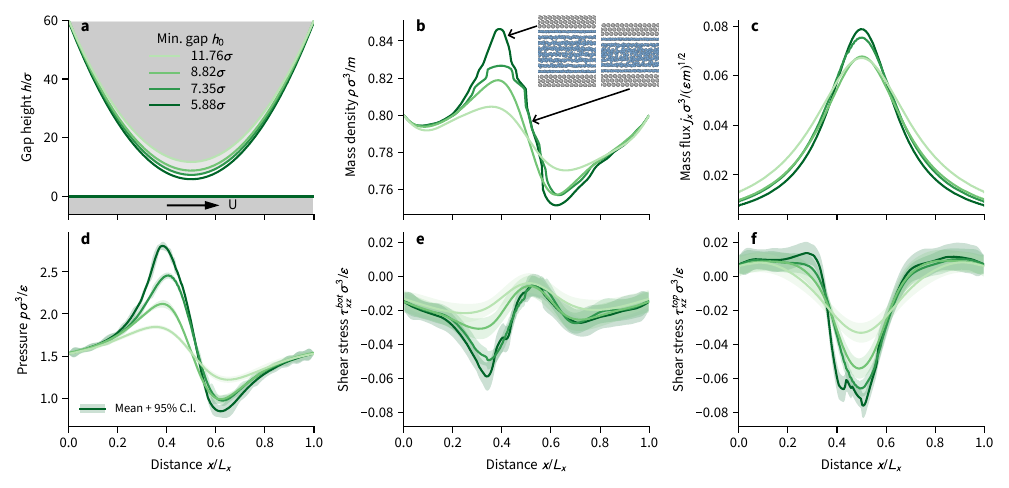}
\caption{%
\textbf{Converged solution and stress profiles for the parabolic slider geometry with different minimum gap height.}
\textbf{a} One-dimensional parabolic slider profile with four different minimum gap heights $h_0=[11.76,\,8.82,\,7.35,\,5.88]\,\sigma$ and constant sliding velocity $U=0.12\velunit$.
The abscissa is scaled by the total length of the system $L_x=1470\sigma$.
\textbf{b} Mass density profile along the converging diverging channel for four gap heights.
Snapshots of the molecular dynamics simulations point to their approximate location and highlight layering transitions as a function of density and gap height.
\textbf{c} Momentum density profiles.
\textbf{d} Pressure profiles with GP posterior mean as a solid line and the shaded area illustrates the 95\% confidence interval based on the posterior variance.
\textbf{e} Shear stress profiles at the bottom wall as predicted by the GP regression model. Mean and confidence interval are illustrated as in \textbf{d}.
\textbf{f} Same as \textbf{e} but for the shear stress at the upper wall.
}
\label{fig:pslider_compare}
\end{figure*}

\subhead{Confinement effects}
To highlight the importance of atomistic effects in boundary lubrication, we simulated a one-dimensional parabolic slider (Fig.~\ref{fig:pslider_compare}a)---a typical benchmark problem in the lubrication literature---and subsequently lowered the minimum size of the constriction towards molecular confinement.
The gap height profile is given by $h(x)=h_0 + 4 (h_1-h_0)/L_x^2 (x - L_x/2)^2$, where $L_x=1470\sigma$ is the length of the domain, and $h_1$ and $h_0$ are the maximum and minimum gap heights, respectively.
Note that here and in the following, all quantities are given in terms of atomic mass ($m$), as well as the length scale ($\sigma$) and the energy parameter ($\epsilon$) of the Lennard-Jones interatomic potential (Eq.~\eqref{eq:lj}) used in the MD simulations (see Methods).
We kept the maximum gap height constant, $h_1=60\sigma$, and considered four different minimum gap heights $h_0=[11.76,\,8.82,\,7.35,\,5.88]\,\sigma$.
The lubricant flow is induced by the flat lower wall, moving with a constant velocity $U=0.12\,\velunit$ in the positive $x$ direction.
All simulations started from constant initial conditions with density $\rho=0.8\densunit$ and flux $j_x=U\rho/2$, the latter corresponding to the gap-averaged Couette mass flow rate.
Dirichlet boundary conditions fixed the mass density at the domain boundaries, with vanishing mass flux gradient (i.e., Neumann BCs for $j_x$).

Figure~\ref{fig:pslider_compare}b shows the converged mass density profile for the four minimum gap heights.
The mass density profiles are characterized by a density peak in front of the constriction, which drops to a level below the ambient density in the diverging part of the profile.
The deviations from the ambient density increase for narrower gaps.
The density profiles for the two larger gap heights, $h_0=11.76\sigma$ and $h_0=8.82\sigma$, appear smooth and look similar to the profiles one would expect from a continuum calculation.
Upon further reduction of the minimum gap, however, discontinuities appear in the density profiles, particularly in regions of high pressure and narrow gaps.
The inset in \ref{fig:pslider_compare}b shows two snapshots of MD simulations under conditions indicated by arrows.
At large densities and small gap heights, the confined fluid forms layers that span the whole gap.
Steps in the density profile indicate transitions between states with different number of layers.

The momentum density profiles shown in Fig.~\ref{fig:pslider_compare}c are approximately symmetric about the center of the domain, with their maxima at the point of the smallest constriction due to momentum conservation.
The gap-averaged mass flux for a pure Couette flow at ambient conditions is $\rho U / 2 = 0.048\fluxunit$.
Deviations from this average flow rate occur due to superimposed pressure-driven flow, enhancing the flow rate in regions of negative pressure gradients, and lowering it in regions of positive pressure gradients.

Fig.~\ref{fig:pslider_compare}d illustrates the GP posterior prediction for the pressure, where the mean is given by the solid lines and the shaded area represents the prediction uncertainty through a 95\% confidence interval.
Here and in the following, when discussing uncertainty in the GP predictions, we exclusively plot and consider confidence intervals based on the GP posterior variance without consideration of the signal noise $\sigma_\mathrm{n}$, i.e., epistemic uncertainty.
The signal noise is intrinsic to the data acquisitions method, here our MD simulations, and depends on the system size and the length of the sampling interval (i.e., aleatoric uncertainty).
Hence, it does not tell much about the predictive ability of the model, but helps to make better predictions since it acts as a regularizer and avoids overfitting.

Pressure excursions in front of the constriction are inversely correlated with the minimum gap height, similar to the mass density profiles.
Pressure maxima coincide with density maxima in front of the constriction and reach up to $2.8\stressunit$, almost twice the ambient pressure.
Behind the constriction, the pressure drops to approximately two thirds of the ambient pressure for the smallest gap, where it also shows the largest uncertainty.
Figure~\ref{fig:pslider_compare}e and f show the GP posterior prediction for the shear stress at the bottom and top wall, respectively.
At the bottom wall, the magnitude of the shear stress is largest at points with large positive pressure gradients, i.e., where pressure-driven flow opposes the shear flow induced by the walls.
A negative pressure gradient with pressure-driven flow in the same direction as the wall movement reduces the shear stress on the bottom wall.
The opposite effect can be observed for the shear stress on the top wall, taking its maximum absolute value in the center of the domain, where the mass flux is largest.
Again, this qualitative behavior is expected from hydrodynamics, but for the two smallest gap heights, anomalies similar to those observed in the mass density are present.
Both stress profiles of the bottom and the top wall show discontinuities and an overall more rugged shape for the two smallest gap heights when compared to the wider gaps.
For instance, in Fig.~\ref{fig:pslider_compare}f the shear stress in the center of the domain for $h_0=5.88\sigma$ clearly deviates from the overall trend seen at the three larger gap heights, indicating that a critical stress has been reached that changes how shear is accommodated by the system.
Integrating normal and shear stress profiles over the whole domain allows calculating a hypothetical friction coefficient.
Here, friction increases by 40\% going from the largest to the smallest gap height.

\begin{figure*}[!ht]
\centering
\includegraphics[width=\textwidth]{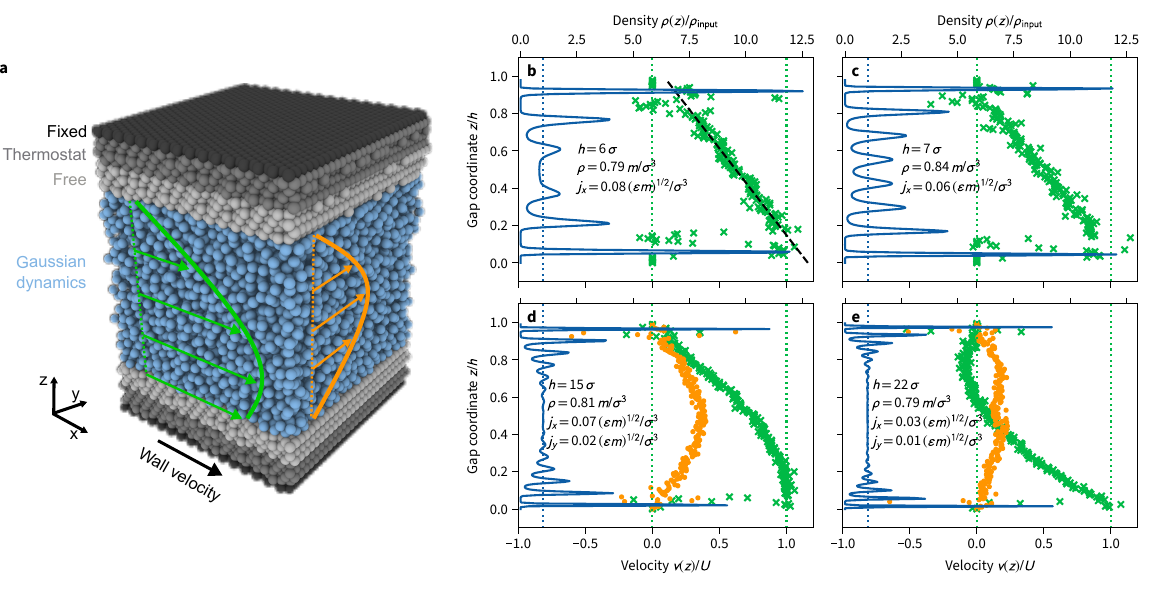}
\caption{%
\textbf{Molecular dynamics (MD) setup and characteristic profiles computed from MD.}
\textbf{a} Slab geometry for the simulation of combined Couette and Poiseuille flow in narrow gaps using nonequilibrium MD simulations.
Interactions between the fluid (blue) and wall (gray/black) atoms are governed by a Lennard Jones (LJ) potential.
The outermost layers (in $z$) of the walls are frozen to fix the gap between the walls and apply shearing at constant velocity.
The central layers of the wall slabs mimic the connection to a heat bath using a Langevin thermostat, keeping the temperature in the system approximately constant.
Pressure-driven flow is enforced by the Gaussian dynamics framework, which maintains a constant mass flow rate in both $x$ and $y$ direction.
\textbf{b} Mass density (upper x-axis) and velocity profile (lower x-axis) for a training simulation in a narrow gap.
Symbols illustrate time averages of atomic velocities in bins distributed across the $z$ coordinate. 
The dashed black line is a linear fit to the velocity data.
\textbf{c} Same as \textbf{b} but at larger density and gap height, but lower mass flux.
\textbf{d} and \textbf{e} Density and velocity profiles at larger gap heights and with nonzero mass flux in the $y$-direction ($v_y$ shown as orange disks).
The configuration is \textbf{d} is located at a point with negative pressure gradient ($v_x(z)$ convex), whereas the configuration in \textbf{e} experiences a positive pressure gradient ($v_x(z)$ concave).}
\label{fig:md_setup}
\end{figure*}

Deviations from the smooth profiles expected from a continuum viewpoint can be explained by atomistic effects.
Figure~\ref{fig:md_setup}a shows the generic MD simulation setup, that generates a combined Poisueille-Couette flow (see Methods), acting as a representative volume element of the local flow state.
The input variables gap height, mass density, and mass flux in the two lateral directions (assuming constant sliding velocity) lead to a diverse collection of flow states, which can be characterized by density and velocity profiles across the gap ($z$-coordinate).
In Fig.~\ref{fig:md_setup}b we show MD simulation results for a flow state close to the point of minimum constriction in Fig.~\ref{fig:pslider_compare} (right inset in b).
The density profile shows the typical layering effect of highly confined fluids, with six layers across the full gap height with $h=6\sigma$.
The outermost fluid layers are more than 12 times denser than the input fluid density $\rho=0.79\densunit$.
Despite the high mass flow rate $j_x=0.08\fluxunit$, a deviation from the linear Couette profile due to the pressure gradient at the point of constriction is not visible.
However, when extrapolating the linear velocity profile to the walls, small deviations from the no-slip assumption become visible.
Thus, fluid-wall slip helps to accommodate the high flow rate rather than a Poiseuille-like velocity profile.
The profiles shown in Fig.~\ref{fig:md_setup}c correspond to a flow state close to the point of maximum pressure/density for the smallest constriction in Fig.~\ref{fig:pslider_compare} (left inset in b).
Although with $h=7\sigma$ only slightly wider, eight fluid layers form within the gap, due to the elevated mass density ($\rho=0.84\densunit$).
At a flow rate of $j_x=0.06\fluxunit$, no slippage occurs between the outermost fluid layers and the walls.

The one-dimensional parabolic slider is an idealization, neglecting the flow in the direction perpendicular to the sliding velocity.
By controlling the flow rate in the perpendicular direction, and measuring the wall shear stress $\tau_{yz}$, we train a separate GP model, which allows us to study realistic gap profiles $h(x, y)$ with our multiscale solver.
Next to the density profiles, Fig.~\ref{fig:md_setup}d and e highlight flow profiles in $x$ and $y$ direction in green and blue, respectively.
At these larger gap heights ($h=[15\sigma,\,22\sigma]$), layering effect are still visible close to the walls, but a homogeneous region without density oscillations develops in the center of the gap, reaching the target densities around $0.8\densunit$.
In sliding direction, mass flow rates above (Fig.~\ref{fig:md_setup}d) or below (Fig.~\ref{fig:md_setup}e) the Couette flow rate lead to the bow-out out of the velocity profiles in the positive or negative $x$-direction, respectively.
Pressure gradients perpendicular to the sliding direction lead to Poiseuille-like profiles.

\subhead{Building a training database}
Our data-driven framework for boundary lubrication relies on an adequate training database incorporating results from MD simulations that represent the local stress states in the confined fluid.
The constitutive relations for the confined fluids are given by GP regression models that interpolate between those stress states measured on the atomistic scale.
A natural choice to generate training data are sampling strategies that optimize space filling, such as Sobol sequences or Latin hypercube sampling.
We used the latter to build a small database of MD simulations close to the initial conditions as a starting point for our multiscale simulations.
Then, the MD database was incrementally augmented by an active learning algorithm during individual simulation runs.
At each time step, we ensure that the maximum variance of the GP posterior distribution for output $Y$ is lower than a pre-defined uncertainty tolerance $\sigma_{\mathrm{t}, Y}^2$, with $\sigma_{\mathrm{t}, Y}=\max(\sigma_{\mathrm{t},Y}^0, \alpha \Delta Y)$.
We chose a combination of absolute and relative tolerances, $\sigma_{\mathrm{t}, Y}=\max(\sigma_{\mathrm{t},Y}^0, \alpha \Delta Y)$, with relative error $\alpha$ w.r.t the maximum difference in the predicted quantity $\Delta Y = \max(Y) - \min(Y)$, and lower bound $\sigma_{\mathrm{t},Y}^0$ for reasonable tolerance levels during start-up from constant initial conditions.
If this is not the case, we run additional MD simulations at the point of maximum variance for this time step.
After convergence to a steady state, the so obtained database of MD runs can be used as a starting point for new multiscale simulations for different geometries and/or with a lower uncertainty tolerance.

\begin{figure*}[!ht]
\centering
\includegraphics[width=\textwidth]{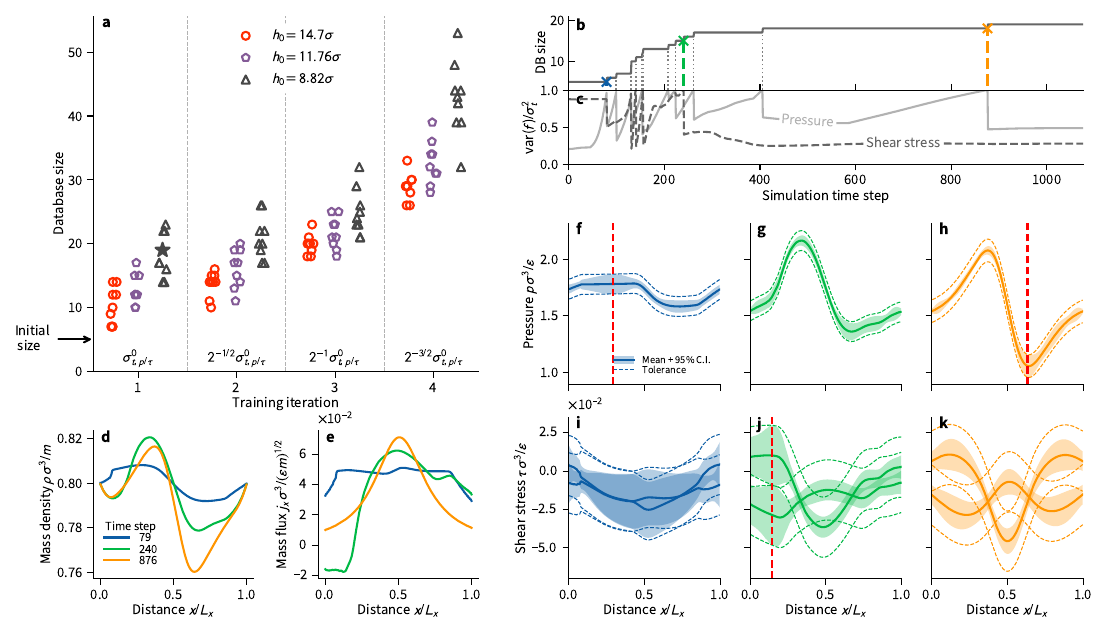}
\caption{%
    \textbf{Data acquisition during training for the one-dimensional parabolic slider.}
    \textbf{a} Size of the molecular dynamics (MD) database after convergence for four training iterations, with subsequently reduced tolerance level and for three gap heights.
    For each gap height, the final database of an iteration served as a starting point for the next finer tolerance level.
    \textbf{b} Growth of the molecular dynamics (MD) database during the first iteration of a training sequence with large uncertainty tolerance $E_{0,p/\tau}$ and minimum gap height $h_0=8.82\sigma$ (highlighted with a star in \textbf{a}).
    The simulation starts with an initial training database of five MD simulations, and dotted vertical lines indicate time steps where the tolerance limit was hit by either of the GP models which triggered new MD runs (shown as maximum variance of the Gaussian process (GP) models normalized by the uncertainty tolerance $\sigma_\mathrm{t}$ in \textbf{c}).
    Three time steps (79, 240, 876) were chosen and marked with blue, green, and orange symbols, respectively. 
    Panels \textbf{d} and \textbf{e} show snapshots of the solution at the highlighted time steps for mass and momentum density (mass flux), respectively.
    Panels \textbf{f}-\textbf{h} and \textbf{i}-\textbf{k} show the prediction of the GP models at the three time instances for pressure and shear stress, respectively, where the shaded region indicates the 95\% confidence interval given the posterior variance around the posterior mean (solid lines).
    The dashed lines highlight the extent of the tolerance band around the mean, and the vertical red dashed line in \textbf{f}, \textbf{j}, and \textbf{h} indicates the physical location, where the tolerance was hit.}
\label{fig:pslider_evolution}
\end{figure*}

Figure~\ref{fig:pslider_evolution}a shows the final size of the MD database for simulations with the parabolic slider geometry for three different minimum gap heights $h_0=[14.70\sigma,\,11.76\sigma,\,8.82\sigma]$.
For each geometry, we subsequently lowered the lower bound of the uncertainty tolerance between individual runs (by halving the variance), starting from $(\sigma^0_{\mathrm{t},p})^2=2\times 10^{-3} \stressunit$ and $(\sigma^0_{\mathrm{t},\tau})^2=1\times 10^{-4} \stressunit$ for pressure and shear stress GP, respectively.
In total, we performed four iterations, i.e., reducing the allowable width of the confidence interval by a factor of $4\sqrt{2}$, while the relative tolerance was kept constant ($\alpha=0.05$).
The database size required for convergence grows with decreasing minimum gap height at constant tolerance level, as well as with decreasing uncertainty tolerance at constant gap height.
Although the size of the final database differs for a sample of training runs ($N=10$) with randomized initial conditions, reproducible results could be obtained with moderate amounts of MD runs.
Once a sufficiently large MD database exists, further refinement of the multiscale solution can be obtained by reducing the relative error $\alpha$, which we did for the two smallest gap heights in Fig.~\ref{fig:pslider_compare}.

Figure~\ref{fig:pslider_evolution}b shows the growth of the MD database for a single simulation of the parabolic slider geometry with $h_0=8.82\sigma$ at the highest uncertainty tolerance (highlighted with a star in Fig.~\ref{fig:pslider_evolution}a).
After a few simulation time steps based on the initial database with five training points, the pressure uncertainty exceeds the tolerance (see Fig.~\ref{fig:pslider_evolution}c) for the first time.
Further additions to the database follow as the simulation proceeds, with eight and six runs triggered by the pressure and shear stress GPs, respectively, leading to a total number of 19 required MD runs.

Next to the first addition to the MD database at time step 79, we highlight the training at time steps 240 and 876 with blue, green, and orange markers.
The colored lines in Fig.~\ref{fig:pslider_evolution}d-k correspond to these time steps.
Figure~\ref{fig:pslider_evolution}d and e show the continuum field variables, mass and momentum density, at the three selected time steps, while Fig.~\ref{fig:pslider_evolution}f-h and Fig.~\ref{fig:pslider_evolution}i-k, show the GP prediction for pressure and shear stress respectively.
In the GP plots, shaded areas highlight the 95\% confidence interval and dashed lines around the confidence band mark the uncertainty tolerance.
The red dashed vertical lines in Fig.~\ref{fig:pslider_evolution}f, h, and j mark the locations, where the uncertainty exceeded the tolerance, which triggered new simulations.
Note that the solution and stress profiles at time step 876 are already close to the converged profiles, but it took more steps to reach a steady state without additionally required MD runs.

\subhead{Multi-asperity and slip-patterned surfaces}
Real world lubrication problems can rarely be reduced to one-dimensional models.
We performed an equivalent training procedure also for two-dimensional problems (see Fig.~\ref{fig:asperity_evolution}), using a gap height profile that mimics a single asperity in a multi-asperity contact and varying minimum gap heights.
The two-dimensional framework uses an additional GP surrogate model for the wall shear stress perpendicular to the sliding velocity.
Yet, the size and the growth of the MD database upon refinement of the tolerance levels is similar to the 1D case (see Fig.~\ref{fig:asperity_training}).

\begin{figure*}
    \centering
    \includegraphics[width=\linewidth]{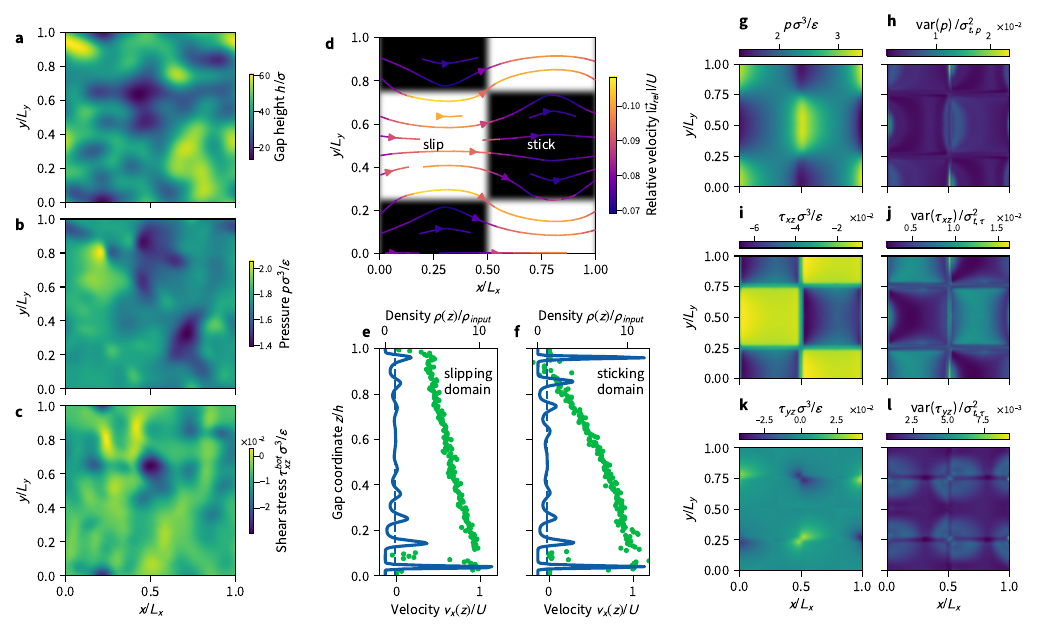}
    \caption{%
    \textbf{Multi-asperity and slip-patterned surfaces.}
    \textbf{a} Self-affine gap height topography for a rough lubricated contact with sliding velocity $U=0.12\velunit$ in the positive $x$-direction.
    \textbf{b} Steady state pressure profile as given by the mean of the Gaussian process (GP) posterior distribution.
    \textbf{c} Shear stress profile ($\tau_{xz}$) at the bottom wall.
    \textbf{d} Checkerboard stick-slip pattern on the surface of the top wall of an otherwise flat channel. Atomistic surfaces in the black area have $\epsilon_\mathsf{wf, slip} = 0.05\epsilon_\mathsf{wf, stick}$.
    Sliding occurs again in positive $x$ direction, leading to the illustrated flow field relative to the lower, moving wall.
    \textbf{e} Density and velocity profile across the gap coordinate for a molecular dynamics (MD) simulation in the sticking domain ($\rho=0.77\densunit$, $j_x=0.074\fluxunit$).
    \textbf{f} Density and velocity profile across the gap coordinate for an MD simulation in the slipping domain ($\rho=0.79\densunit$, $j_x=0.064\fluxunit$) with significant wall slip at the lower surface. Density layering is less pronounced at the slip wall.
    \textbf{g} Steady state pressure profile as given by the mean of the Gaussian process (GP) posterior distribution.
    \textbf{h} Variance of the pressure GP at steady state, normalized by the uncertainty tolerance $\sigma_{\mathsf{t}, p}$.
    \textbf{i} Steady state shear stress profile at the lower wall ($\tau_{xz}$) as given by the mean of the GP posterior distribution.
    \textbf{j} Variance of the shear stress GP at steady state, normalized by the uncertainty tolerance $\sigma_{\mathsf{t}, \tau}$.
    \textbf{k} Same as \textbf{h} but for $\tau_{yz}$.
    \textbf{l} Same as \textbf{i} but for $\tau_{yz}$.
    }
    \label{fig:2d_examples}
\end{figure*}

We used the training database generated with the single asperity gap profile for simulations of multi-asperity gaps arising in the lubrication of rough surfaces.
Figure~\ref{fig:2d_examples}a shows the gap height profile for such a system with mean height $\langle h \rangle=38\sigma$ and self-affine surface topography with root-mean-square roughness $\langle(h - \langle h \rangle)^2\rangle^{1/2}=6\sigma$ generated from a power spectral density with Hurst exponent $H=0.8$ for roughness wavelengths $400\leq\lambda/\sigma\leq1400$.
The side length $L_x=L_y=2940\sigma$ of the square domain was discretized using 256$\times$256 finite volume cells with periodic boundary conditions in both directions.
Although trained on a much simpler geometry, only eleven additional MD training runs were requested by the active learning algorithm during the simulation (using the same tolerance level as in the last step of the training).
The steady state pressure map for the two-dimensional domain in Fig.~\ref{fig:2d_examples}b shows pressure excursions in front of constrictions in analogy to the one-dimensional example.
Shear stress maps can be obtained in the same way from the GP shear stress model, as shown in Fig.~\ref{fig:2d_examples}c for $\tau_{xz}$ on the bottom wall.
Tangential and normal forces, i.e., the integrals over the predicted stress profiles, determine the macroscopic friction coefficient.
Thus, in principle, our framework allows for systematic characterization of frictional properties with atomistic resolution.

Pressure excursions due to height variation in lubricated gaps dominate the load bearing capacity of the fluid film.
Next to the gap height, surface properties may also vary along the lateral dimensions of a lubricated gap.
It has been shown that heterogeneous surface properties may also contribute to the load bearing capacity~\citep{savio2016_boundary}.
We used this fact to highlight the flexibility and extensibility of our framework by simulating the shear-driven flow between two parallel walls, where the upper wall surface has a checkerboard pattern, as shown in Fig.~\ref{fig:2d_examples}d.
Neighboring square patches on the checkerboard have alternating surface energies, which manifest themselves in a different effective wall slip between the outermost fluid layer and the solid.
In practice, the surface energy has been adjusted through the wall fluid interaction ($\epsilon_\mathrm{wf}^\mathrm{slip}=\kappa \epsilon_\mathrm{wf}^\mathrm{stick}$) in the MD simulations, with scale factor $0\leq\kappa<1$.
Since the height is constant, it does not need to be considered as an input to the GP models.
Instead, we used the scale factor $\kappa$ as a continuous input to both the pressure and shear stress GP models.
We used $\kappa=0.05$ for slippery patches and $\kappa=1$ for sticky patches on the upper wall, and interpolated smoothly between those values across the domain boundaries.
Following a training procedure akin to the previous examples, we computed the steady state solution for the heterogeneous channel, which is illustrated by the streamlines in Fig.~\ref{fig:2d_examples}d.
The streamlines illustrate the gap-averaged velocity field relative to the ideal Couette flow without slip and pressure gradient, i.e., $\vec{u}_\mathrm{rel}=(j_x/\rho - U/2, j_y/\rho)^\top$.
The effect of the modified fluid-wall interface becomes evident in the density and velocity profiles of the MD calculations.
Fig.~\ref{fig:2d_examples}e and f show the density and velocity profiles ($v_x$) of MD training simulations in the slipping and sticking domain, respectively.
While the profiles in the pure sticking domain show the typical density layering and no-slip velocity profiles, the weaker interaction between the upper wall and the fluid in the slipping domains lead to significantly reduced density oscillations and an increased wall slip.
The predicted pressure profile and uncertainty at steady state as given by the GP posterior mean and variance, respectively, are shown in Fig.~\ref{fig:2d_examples}g and h.
The heterogenous surface properties lead to a pressure profile with maxima and minima at the transition from slipping to sticking domains (in sliding direction) and vice versa, respectively.
This is a direct effect of momentum conservation, which is altered due to the change in the wall boundary conditions.
The uncertainty of the pressure prediction is largest at both extrema, but relatively low compared to the selected threshold for active learning.
The magnitude of the shear stress profile $\tau_{xz}$ at the upper wall is lower on the slipping patches than on the sticking patches (see Fig.~\ref{fig:2d_examples}i), which is expected due to the reduced interaction.
The shear stress on the top wall in the direction perpendicular to the flow is shown in Fig.~\ref{fig:2d_examples}k, with maxima and minima located at the corners, and has low uncertainties (Fig.~\ref{fig:2d_examples}j and l).

\section*{Discussion}

Our multiscale coupling scheme enables to include atomistic effects into continuum representations of lubricant flow.
The stress surrogate models capture confinement effects such as fluid layering or wall slip and transfer them to the continuum solver.
Thus, we were able to solve lubrication problems on length scales which are impractical or even unfeasible with all-atom simulations.
Interpolation of the surrogate models combined with active learning relaxes the enslavement to the time step of the atomistic scale, requiring  only a few representative MD simulations to reach a steady state (see Fig.~\ref{fig:pslider_evolution}a).
Lowering the uncertainty tolerance of the active learning scheme allows fine-tuning the resolution of the macroscopic solution, leading to a trade-off between accuracy and computational cost.
Although we investigated only a simple model fluid, we believe that our framework is directly transferable to more complex fluid models to describe the behavior of real lubricants.

Our multiscale solver measures stress tensor components directly, thereby avoiding the detour of finding suitable parametrizations for, e.g., viscosity or slip length.
Machine learning surrogates for the stress tensor~\citep{frankel2020_tensor,lennon2023_scientific} are often designed to fulfill thermodynamic constraints, such as objectivity or isotropy.
The reduced order model of the lubrication problem circumvents such considerations, since we do not need to measure the full 3D stress tensor.
In fact, we deliberately designed our method to work in cases where continuum assumptions break, such as for strong confinement.
Nevertheless, thermodynamic constraints might confine the allowable range of hyperparameters \citep{sharma2024_learning}.

The choice of input variables controls the level of detail at which atomistic effects are resolved.
For instance, if we had chosen the applied strain rate $\dot{\gamma}=U/h$ as an input variable, instead of wall velocity $U$ and gap height $h$ independently, density layering at small gap heights (e.g., Fig.~\ref{fig:md_setup}b-c) would have been homogenized in the constitutive surrogate model.
The flexibility of the GP surrogate models were underpinned by the extension to two-dimensional problems, i.e., going from three to four input dimensions (Fig.~\ref{fig:2d_examples}a-c), or by considering structured surfaces by replacing one of the input variables (Fig.~\ref{fig:2d_examples}d-l).

The computational cost for inference in GP regression grows with the cube of the number of training points (MD simulations).
Although the MD will likely stay the bottleneck of our method, more complex models (e.g., with many input dimensions) might require us to resort to sparse GPs, which build the covariance matrix with only a subset of the training data~\citep{rasmussen2006_gaussian}.
So far, we have used ``vanilla'' GP models, which worked already quite well for nontrivial problems but may not always be an optimal choice.
For instance, when approaching severe confinement as in Fig.~\ref{fig:pslider_compare}, rapidly changing outputs dominate the covariance everywhere in the input space, which requires more training also in regions where not much happens.
Such, non-stationary behavior can not be modelled with covariance kernels that only depend on relative distances.
So-called deep GPs~\citep{damianou2013_deep} borrow ideas from deep neural networks, and warp the input space by putting a GP prior on the inputs.
Several of these nonlinear input transformations can be stacked on top of each other, each one acting as a latent variable, which are then used as inputs to common stationary kernels~\citep{damianou2013_deep,sauer2023_active}.

Previous multiscale schemes for thin film or lubricated flow have achieved coupling through an exchange of flux and pressure gradient variables between the atomistic and the continuum representation~\citep{chandramoorthy2018_solving,borg2013_fluid, stephenson2018_accelerating}.
Usually, a constant force corresponding to a continuum pressure gradient is applied to all atoms in an atomistic representative volume element, and the resulting mass flux is measured.
This assumes linear response in order to be able to consider Couette and Poiseuille contributions separately~\citep{chandramoorthy2018_solving}, and requires similar interfacial conditions at both top and bottom wall.
However, effects such as fluid wall slip at opposing walls of a sheared fluid film can differ substantially due to different local strain rates (e.g., Fig.~\ref{fig:md_setup}d) or different wall constituents (Fig.~\ref{fig:2d_examples}d).
In contrast, we control the densities of conserved variables on the atomistic level according to our continuum solution and measure the corresponding fluxes in the system.
Hence, we control the mass flux in a thermodynamically consistent way by applying a fluctuating (in time) force to fluid atoms according to the Gauss' principle of least constraint~\citep{strong2017_dynamics}, and measure stress tensor components which manifest themselves as wall traction.\footnote{Note that mass flux occurs both as a controlled variable (in the momentum balance) and as a flux (in the mass balance). While stress describes the flux of momentum and depends on the constituents, mass conservation is always exactly fulfilled, i.e., we do not need to measure a ''resulting`` mass flux.}
This framework can be extended by including more conservation laws, such as the conservation of energy, which would require the measurement of heat fluxes in the system.
Here, we assume that any heat generated due to viscous heating of the thin film is immediately conducted away to the heat bath, which justifies the isothermal assumption.

One of the most important quantities of interests in tribology is the coefficient of friction, i.e., the ratio of friction forces to the applied normal load.
When predictions of multiscale models such as ours will eventually be compared to experiments, our choice of coupling variables allows a direct propagation of model uncertainties in the pressure and shear stress to the coefficient of friction.
This avoids expensive sampling techniques for error propagation and helps to interpret simulation results, particularly in lubrication regimes under extreme conditions.

In the future, the MD training database may not only be limited to data that we computed on our own, but may profit from the vast computational expertise and efforts of fellow tribologists.
This would require standardized output formats, metadata, and public repositories following the FAIR~\citep{wilkinson2016_fair} principles, a topic currently under active development in tribology~\citep{garabedian2022_generating} and the entire materials science community~\citep{bayerlein2024_concepts}.
Within our framework, we implemented data management with \textit{dtool}~\citep{hormann2024_dtool}, which might be extended to interact with existing infrastructure.

\section*{Materials and Methods}

\subhead{Continuum description of thin film flows}
We denote the density field of a conserved variable with $\mathbf{q} \equiv \mathbf{q}(\vec{r},t)$.
Thus, the time evolution of $\mathbf{q}$ is given by the continuity equation
\begin{equation}\label{eq:conservation}
    \frac{\partial \mathbf{q}}{\partial t} +
    \frac{\partial \mathbf{f}_x}{\partial x} +
    \frac{\partial \mathbf{f}_y}{\partial y} +
    \frac{\partial \mathbf{f}_z}{\partial z} = 0,
\end{equation}
where $\mathbf{f}_i \equiv \mathbf{f}_i(\vec{r},t)$ describes the corresponding flux in the Cartesian direction $i\in [x, y, z]$.
Note that here and in the following, bold symbols (e.g., $\mathbf{q}$, $\mathbf{f}$) indicate vectors of arbitrary length representing a collection of variables while arrows (e.g., $\vec{r}$, $\vec{k}$) indicate Cartesian 3-vectors.
Equation~\eqref{eq:conservation} generally holds for conserved quantities, but here, we focus on the conservation of mass and linear momentum, given by their densities $\rho$ and $\vec{j}\equiv \rho\vec{u}$, respectively, where $\vec{u}$ describes a velocity field.
Mass and momentum flux vectors are then given by $\mathbf{f}_i=(\vec{j}\cdot\hat{\vec{e}}_i, [(\vec{j}\otimes\vec{j})/\rho + p \underline{1} -\underline{\tau}] \hat{\vec{e}}_i)^\top$, where $\underline{\tau}$ is the viscous stress tensor, $p$ is the pressure, which for an isothermal compressible fluid is given by an equation of state $p(\rho)$, and $\underline{1}$ and $\hat{\vec{e}}_i$ denote the unit matrix and vector in the Cartesian basis, respectively.
Expressing $\underline{\tau}$ as a linear function of the symmetric gradient of $\vec{u}$ (Newtonian fluid), yields the well-known Navier-Stokes equation.
For thin film flows, proper dimensional scaling leads to dimensionality reduced forms of the Navier-Stokes equation, which can be readily solved.

Here, we seek a lower dimensional description of the thin film flow problem, which should be agnostic to the particular form of the constitutive laws $\underline{\tau}(\rho, \vec{j})$ and $p(\rho)$.
We briefly recap the main ideas of this approach, but refer to Ref.~\citep{holey2022_heightaveraged} for a more detailed derivation and validation.
We want to solve Eq.~\eqref{eq:conservation} in a thin gap bounded by two walls, given by $\Omega=\{\vec{r}\in \mathbb{R}^3: [0, L_x] \times [0, L_y] \times [h_1, h_2] \}$, where $h_1\equiv h_1(x, y, t)$ and $h_2\equiv h_2(x, y, t)$ describe the topography of the lower and upper surface, respectively, and the gap profile $h(x, y, t) = h_2(x, y, t) - h_1(x, y, t)$ is much smaller than the lateral dimensions $L_x$ and $L_y$.
Thus, we integrate Eq.~\eqref{eq:conservation} across the small dimension $z$, i.e.,
\begin{equation}\label{eq:conservation_integral}
    \int_{h_1(x,y,t)}^{h_2(x,y,t)} \left( \frac{\partial \mathbf{q}}{\partial t} +
    \frac{\partial \mathbf{f}_x}{\partial x} +
    \frac{\partial \mathbf{f}_y}{\partial y} +
    \frac{\partial \mathbf{f}_z}{\partial z}\right) dz = \mathbf{0}.
\end{equation}
considering the rules for differentiation under the integral sign, which leads to 
\begin{equation}\label{eq:avg_balance}
    \frac{\partial \bar{\mathbf{q}}}{\partial t}  + \frac{\partial \bar{\mathbf{f}}_x}{\partial x} + \frac{\partial \bar{\mathbf{f}}_y}{\partial y} + \mathbf{s} = \mathbf{0},
\end{equation}
where overbars denote gap-averaged fields (e.g., $\bar{\mathbf{q}}=h^{-1}\int_{h_1}^{h_2}\mathbf{q}dz $), and $\mathbf{s}$ is a source term that arises from the average
\begin{equation}
\begin{split}
\bm{s}
&= \frac{1}{h}\Biggl[
 \frac{\partial h_2}{\partial x}(\bar{\bm{f}}_x - \bm{f}_x\rvert_{z=h_2})
-\frac{\partial h_1}{\partial x}(\bar{\bm{f}}_x - \bm{f}_x\rvert_{z=h_1}) 
+\frac{\partial h_2}{\partial y}(\bar{\bm{f}}_y - \bm{f}_y\rvert_{z=h_2})
-\frac{\partial h_1}{\partial y}(\bar{\bm{f}}_y - \bm{f}_y\rvert_{z=h_1})  \\
&\phantom{=\frac{1}{h}\Biggl[}
-\frac{\mathrm{d} h_2}{\mathrm{d} t}(\bar{\bm{q}} - \bm{q}\rvert_{z=h_2})
+\frac{\mathrm{d} h_1}{\mathrm{d} t}(\bar{\bm{q}} - \bm{q}\rvert_{z=h_1})  
+ \bm{f}_z\rvert_{z=h_2} - \bm{f}_z\rvert_{z=h_1}\Biggr].
\end{split}
\end{equation}
Without loss of generality, we simplify the source term by assuming a flat lower wall $h_1=\mathrm{const.}$ sliding against an upper stationary profile $h_2(x, y)$, leading to
\begin{align}
\bm{s}
&= \frac{1}{h}\left[
 \frac{\partial h_2}{\partial x}(\bar{\bm{f}}_x - \bm{f}_x\rvert_{z=h_2})
+\frac{\partial h_2}{\partial y}(\bar{\bm{f}}_y - \bm{f}_y\rvert_{z=h_2})
+ \bm{f}_z\rvert_{z=h_2} - \bm{f}_z\rvert_{z=h_1}\right].
\end{align}
Hence, the time evolution of $\bar{\mathbf{q}}$ is fully determined by average flux components in $x$ and $y$ direction, as well as unaveraged flux components evaluated at the top and bottom wall.
This approach is similar to conventional thin film descriptions, such as the Reynolds equation for lubrication, with the only difference that we make no assumption about the constitutive relation of the fluid.
Assuming that the density field can be additively split into a gap-averaged and a $z$-dependent part, i.e., $\mathbf{q}(x,y,z,t)=\bar{\mathbf{q}}(x, y, t) + \delta\mathbf{q}(z)$, where $\delta\mathbf{q}(z)$ depends only implicitly on $x$ and $y$ via $\bar{\mathbf{q}}$, we showed that the lubrication problem can be formally split into two subproblems: one so-called macro problem, that describes the time evolution of the gap-averaged fields according to Eq.~\eqref{eq:avg_balance}, and a micro problem that resolves the flow across the gap.
The two subproblems are coupled, since the micro problem determines the relevant fluxes for the time evolution of $\bar{\mathbf{q}}$, which in turn defines the constraints for the micro problem.
We used a finite volume description of the two-dimensional domain, and solve the macro problem using an explicit time integration scheme \citep{maccormack2003_effect}.
Non-equilibrium molecular dynamics (MD) simulations of confined fluids provide the solution of the micro problem, i.e., the fluxes required for closure of the macro problem.

\subhead{Molecular dynamics simulations of confined fluids}
We performed non-equilibrium molecular dynamics simulations of confined fluids between flat atomistic walls and determined the normal and shear stress components in these systems.
Temporal averages of the measured stress constitute the database to train a surrogate model, which is used to close the evolution equation on the macro scale.
In order to provide a proof of concept, we use a simple model system, but extension to more advanced models for realistic lubricants and interfaces should be straightforward.

Hence, all interatomic forces between pairs of fluid atoms, pairs of wall atoms, as well as fluid-wall interactions are governed by the Lennard-Jones interatomic potential
\begin{equation}
\label{eq:lj}
U(r_{ij}) = 4\epsilon\left[\left(\frac{\sigma}{r_{ij}}\right)^{12} - \left(\frac{\sigma}{r_{ij}}\right)^6\right], \; r_{ij} < r_\mathrm{c},
\end{equation}
where $r_{ij}=|\vec{r}_i - \vec{r}_j|$ is the interatomic distance, $r_\mathrm{c}$ is the cutoff radius, and $\epsilon$ and $\sigma$ are the energy and length scale parameters of the pair potential, respectively.
We cut the interatomic potential at $r_\mathrm{c}=2.5\sigma$ and shift energies by $U(r_\mathrm{c})$ to reach zero at the cutoff.
Upper and lower walls consist of atoms arranged in an FCC single crystalline structure with lattice constant $a=1.2\sigma$, which leads to a wall density of $2.31\sigma^{-3}$.
The \{111\} planes are exposed to the fluid and the periodic wall slabs have lateral dimensions which are integer multiples ($n$) of $4a\sqrt{3/2}$ and $7a\sqrt{2}/2$, respectively, leading to a nearly quadratic surface area (we used $n=3$ for most of our simulations, but at least $n>1$).
The walls consist of nine \{111\} layers each, where the outermost wall layers are frozen, a central region of four layers is used to apply a thermostat, and the remaining four layers undergo free dynamics.
We apply the Lorentz-Berthelot mixing rules for the interactions between fluid and wall atoms, where length scale and energy parameters are obtained via arithmetic, $\sigma_\mathrm{wf}=(\sigma_\mathrm{f}+\sigma_\mathrm{w})/2$, and geometric mean, $\epsilon_\mathrm{wf}=(\epsilon_\mathrm{w}\epsilon_\mathrm{f})^{1/2}$, of the individual parameters, respectively.
Note that in our simulations, we set $\sigma_\mathrm{f}$ and $\epsilon_\mathrm{f}$ as well as the mass $m$ of a fluid atom to unity, and the wall parameters are chosen such that the ratio of length scales, masses, and energies resembles that of a model argon/gold interface \citep{heinz2008_accurate}.
For systems with heterogeneous surfaces, we scaled the energy parameter accordingly ($\epsilon_\mathrm{wf}^\mathrm{slip}=\kappa \epsilon_\mathrm{wf}^\mathrm{stick}$), with scale factor $0\leq\kappa<1$.

Initially, we placed two identical wall slabs at a distance of $h+\sigma$ apart in the $z$-direction, and filled the space between them with fluid atoms at random positions.
We apply a Berendsen thermostat to all fluid atoms for $50\,000$ time steps and restrict the maximum distance per step to $0.1\sigma$ to avoid a blow up of the system from overlapping particles due to the random initialization.
Following the initial thermalization, rigid atoms of the lower wall are then displaced with a constant velocity $U$ to impose a shear flow.
We apply a Langevin thermostat to the central region of each wall acting only on the peculiar velocities, with a coupling time of $100\Delta t$, where $\Delta t= 0.005 \tau$ is the integration time step and $\tau=\sqrt{m\sigma/\epsilon}$.
At the same time, we integrate the positions and momenta of the fluid atoms using the Gaussian dynamics scheme \citep{strong2017_dynamics}, which fixes the overall mass flux vector by applying a force correction in the velocity Verlet time integration.
Thus, we constrain the mass flux in $x$- and $y$-direction to the value given by our continuum solution, thereby simulating a combined shear- and pressure-driven flow in the system.
When a steady state is reached (usually after another $50\,000$ time steps), we start sampling of the stress in the MD systems.
We have used a variety of sampling times, which affects the uncertainty of the measured stress averages, but all the results shown here, have been sampled for at least $100\,000$ steps.

We probe the normal and shear stress in the confined fluid under the combined pressure and shear driven flow.
While the global pressure of a homogeneous system in equilibrium is unambiguously defined by the virial theorem, difficulties arise e.g., in non-equilibrium systems, when long-range interactions have to be considered, or, when the stress tensor needs to be resolved locally in a heterogeneous system, such as the confined fluids we investigate here.
We refer to Ref.~\citep{shi2023_perspective} for a recent review of the various aspects of molecular stress tensor computation.
Our coupling framework requires stress tensor components both in a gap-averaged sense and evaluated at the wall fluid interface.
If the wall separation is large enough that interfacial effects do not extend to the center of the fluid film, the former can be obtained by evaluating the virial expression in a region in the center of the channel.
Here, we use a mechanical definition of the stress tensor, which we probe as surface traction at the interface between fluid and wall.
The only relevant gap-averaged stress is the pressure $p$, which we calculate as the mean of the normal stress on the upper and lower wall.

\subhead{Gaussian process regression and active learning}
Our multiscale scheme is based on the propagation of continuous field variables, while the stresses required to close the system are calculated in representative MD simulations.
Computing stresses from MD at every macro time step and at each grid point would lead to many redundant MD runs. 
Instead, we formulate a surrogate model for the stress, which is informed by the MD data.
We use Gaussian processes to describe the relation between input $\mathbf{X}$ and stress output $f$,
\begin{equation}
f(\mathbf{X}) \sim \mathcal{GP}(\mu(\mathbf{X}), k(\mathbf{X}, \mathbf{X}^\ast)),
\end{equation}
where each output describes a multivariate Gaussian normal distribution with mean given by the mean function $\mu(\mathbf{X})$ and covariance given by the kernel $k(\mathbf{X}, \mathbf{X}^\ast)$.
Here, we use a zero mean function and the Matérn3/2 kernel to describe correlations among inputs, 
\begin{equation}
k(\mathbf{X}_i, \mathbf{X}_j) = \sigma^2\left(1+\sqrt{3}d(\mathbf{X}_i, \mathbf{X}_j)\right)\exp\left(-\sqrt{3}d(\mathbf{X}_i, \mathbf{X}_j)\right),
\end{equation}
where $d(\mathbf{X}_i, \mathbf{X}_j)=\sum_{k=1}^{n_d} l_k^{-1}\sqrt{(X_{j,k} - X_{i,k})^2}$ is the scaled distance between inputs $\mathbf{X}_i$ and $\mathbf{X}_j$ and $\sigma^2$ is the kernel variance that describes the uncertainty at distances much larger than the correlation length scales $l_k$.
Furthermore, we assume noisy observations, where the noise is normally distributed with zero mean and variance $\sigma_n^2$.
In Gaussian process regression \citep{rasmussen2006_gaussian}, Bayesian inference allows updating our prior belief by conditioning the predictive distribution on the training data $\{\mathbf{X}^\ast, \mathbf{Y}^\ast\}$.
The posterior mean function is then given by
\begin{equation}
\bar{f} = \mu(\mathbf{X}) + k(\mathbf{X}, \mathbf{X}^\ast) \left[k(\mathbf{X}^\ast, \mathbf{X}^\ast) + \sigma_n^2 \bm{I}\right]^{-1} (\mathbf{Y}^\ast - \mu(\mathbf{X}^\ast)),
\end{equation}
and the covariance is given by
\begin{equation}
\text{cov} (f) = k(\mathbf{X}, \mathbf{X}^\ast) - k(\mathbf{X}, \mathbf{X}^\ast) \left[k(\mathbf{X}^\ast, \mathbf{X}^\ast) + \sigma_n^2 \bm{I}\right]^{-1} k(\mathbf{X}^\ast, \mathbf{X}). 
\end{equation}

Here, predictive variables are the shear stress components $\tau_{xz}$ and $\tau_{yz}$, both evaluated at the top and bottom wall, as well as the gap-average of the normal stress, i.e., the pressure $\bar{p}$.
The continuum solution $\mathbf{q}^n\equiv \mathbf{q}(x, y; t_n)$ at time $t_n$ is used to map from the two-dimensional Cartesian space to the input space of the GP.
Besides the three continuum field variables, boundary conditions such as the constant gap height profile $h(x,y)$, wall properties $\kappa(x,y)$, and the wall sliding speed $U$ can be used as inputs.
In this work, we used $\mathbf{X}=(h, \rho, j_x, j_y)^\top$ for simulations with varying gap height, and $\mathbf{X}=(\kappa, \rho, j_x, j_y)^\top$ for simulations with constant gap but slip-patterned  surface.

In practice, we use three separate models for normal stress (pressure) $p$, shear stress $\tau_{xz}$ and $\tau_{yz}$.
To predict shear stresses at the bottom and top wall simultaneously, we use multi-output GPs assuming no correlation among the outputs but equal correlations among inputs of the same output dimension. 
This corresponds to a sum of separable kernels, leading to a block-diagonal kernel matrix with identical blocks on the diagonal \citep{alvarez2012_kernels}, i.e., the same set of hyperparameters for all dimensions.
The prediction strongly depends on the choice of hyperparameters.
With four length scales and one variance parameter per kernel, this culminates to 15 parameters, which we  optimize by maximizing the logarithmic marginal likelihood 
\begin{equation}
\log\, P(\mathbf{Y}^\ast|\mathbf{X}^\ast) = -\frac{1}{2}\mathbf{Y}^{\ast\top}[k(\mathbf{X}^\ast,\mathbf{X}^\ast) +\sigma_n^2\mathbf{I}]^{-1}\mathbf{Y}^\ast -\frac{1}{2}\log|k(\mathbf{X}^\ast,\mathbf{X}^\ast) +\sigma_n^2\mathbf{I}| + \frac{n}{2} \log 2\pi
\end{equation}
of the GP w.r.t. the hyperparameters.
We use a gradient based minimization of the negative marginal log likelihood and randomly perturb starting configurations to avoid being trapped in local minima.
The initial starting configurations are found via grid search for the length scales and analytical optimization of the corresponding variance~\citep{basak2022_numerical}.
We check at each time step $n$, whether the prediction $\mathbf{Y}_\mathrm{n}= f\,|\,\mathbf{X}_\mathrm{n}, \mathbf{X}^\ast, \mathbf{Y}^\ast$ meets a predefined uncertainty tolerance $\sigma_\mathrm{t}$, i.e., we check that $\operatorname{max} \operatorname{var}(\mathbf{Y}_\mathrm{n}) < \sigma_\mathrm{t}^2$.
If this is not the case, new input variables are chosen at the location where the prediction uncertainty is largest, i.e., $\mathbf{X}_\mathrm{new}=\operatorname{argmax}_{\mathbf{X}_\mathrm{n}} \operatorname{var} (\mathbf{Y}_\mathrm{n})$.

\FloatBarrier

\begin{acknowledgments}
The authors gratefully acknowledge support by the German Research Foundation (DFG) through GRK 2450.
We used \textsc{LAMMPS}~\citep{thompson2022_lammps} for all molecular dynamics simulations and \textsc{GPy}~\citep{gpy2014} as Gaussian process library.
Snapshots of MD simulations in Fig.~\ref{fig:workflow}-\ref{fig:md_setup} have been taken with \textsc{Ovito}~\citep{stukowski2009_visualization}.
Data management has been implemented with \textsc{dtool} and \textsc{dserver}~\citep{hormann2024_dtool}.
We thank Johannes Hörmann for technical assistance and server administration.
\end{acknowledgments}

\appendix

\FloatBarrier

\newpage

\renewcommand\thefigure{S\arabic{figure}}   
\setcounter{figure}{0}    

\section*{Supplementary figures}

\begin{figure*}[!ht]
\centering
\includegraphics[width=\textwidth]{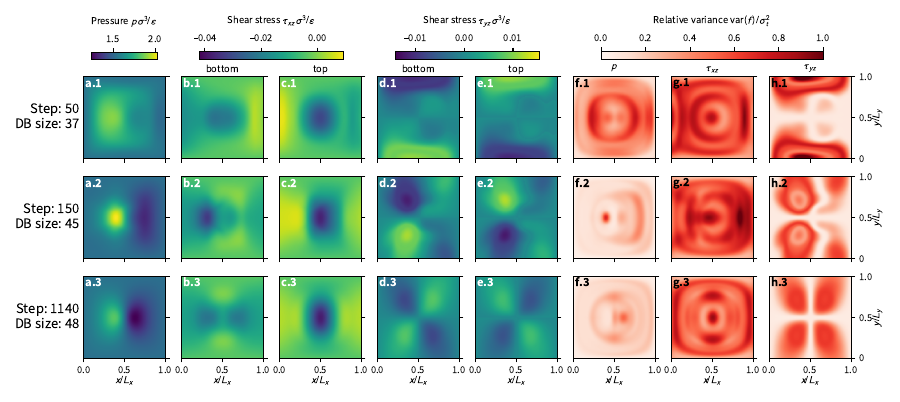}
\caption{%
\textbf{Evolution of pressure and shear stress predictions during a training run on a two-dimensional example problem mimicking the flow around a single asperity.}
Column \textbf{a} illustrates the Gaussian process (GP) posterior mean pressure predictions for three different simulation time steps.
The minimum gap height in the center of the domain is $8.82\sigma$.
The first row corresponds to an initial stage of the simulation, with a prediction based on data from 37 molecular dynamics runs.
The second row corresponds to a time frame one hundred steps later with 45 MD runs, and the third and last row shows the converged solution after 1140 steps and 48 MD runs.
Columns \textbf{b} and \textbf{c} show the shear stress predictions of the GP in the shearing direction $\tau_{xz}$ at the bottom and top wall, respectively, for the three time frames.
Accordingly, the shear stress perpendicular to the shearing direction $\tau_{xz}$ is shown in columns \textbf{d} and \textbf{e}.
The last three columns \textbf{f}, \textbf{g}, and \textbf{h}, show the posterior variance of the GP regression models for pressure and the two shear stress components, respectively.
The variance is normalized by the corresponding uncertainty tolerance which is $E_p=2.5\times10^{-4}$ for the pressure and $E_\tau=1.25\times10^{-5}$ for the shear stress, i.e., the largest refinement level used for the single asperity training calculations.
Hence, dark red regions indicate low predictive confidence and brighter regions correspond to regions with relatively high confidence.
}
\label{fig:asperity_evolution}
\end{figure*}

\begin{figure}[!ht]
    \centering
    \includegraphics{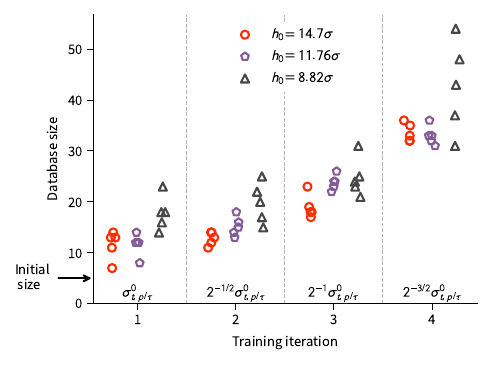}
    \caption{\textbf{Size of the training database for the single asperity geometry.} Same as Fig.~\ref{fig:pslider_evolution}a of the main text but for two-dimensional simulations.}
    \label{fig:asperity_training}
\end{figure}

\end{document}